\documentclass[useAMS,usenatbib,usegraphicx]{mn2e}

\newcommand{\co}     {$^{13}$CO\,(1--0)}
\newcommand{\cvo}    {C$^{18}$O\,(1--0)}
\newcommand{\dvco}   {$^{12}$CO\,(1--0)}
\newcommand{\cs}     {CS\,(2--1)}
\newcommand{\ctrs}   {C$^{34}$S\,(2--1)}
\newcommand{\kms}    {km\,s$^{-1}$}
\newcommand{\offsets}{($\Delta \alpha, \Delta \delta$)}
\newcommand{\vlsr}   {$V_{\rm lsr}$}
\newcommand{\ha}     {H$\alpha$}
\newcommand{\hii}    {H\,{\sc{ii}}}

\title[Star formation around the \hii\ region Sh2-235]{Star formation around the \hii\ region Sh2-235}
\author[M. S. Kirsanova et al.]{M. S. Kirsanova$^{1}$,\thanks{E-mail:
kirsanova@inasan.ru}
A. M. Sobolev$^2$,
M. Thomasson$^3$,
D. S. Wiebe$^1$,
\newauthor L. E. B. Johansson$^3$,
A. F. Seleznev$^2$\\
$^{1}$Institute of Astronomy of the RAS, 119017, Moscow, Russia\\
$^{2}$Ural State University, 620051, Ekaterinburg, Russia\\
$^{3}$Onsala Space Observatory, SE--439 92, Onsala, Sweden}

\begin{document}

\date{Accepted 2008 May 3. Received 2008 April 23; in original form 2008 February 11}

\pagerange{\pageref{firstpage}--\pageref{lastpage}} \pubyear{2008}

\maketitle

\label{firstpage}

\begin{abstract}
We present a picture of star formation around the \hii\ region Sh2-235 (S235) based upon data on the spatial distribution of young stellar clusters and the distribution and kinematics of molecular gas around S235. We observed \co\ and \cs\ emission toward S235 with the Onsala Space Observatory 20-m telescope and analysed the star density distribution with archival data from the 2MASS survey. Dense molecular gas forms a shell-like structure at the south-eastern part of S235. The young clusters found with 2MASS data are embedded in this shell. The positional relationship of the clusters, the molecular shell and the \hii\ region indicates that expansion of S235 is responsible for the formation of the clusters. The gas distribution in the S235 molecular complex is clumpy, which hampers interpretation exclusively on the basis of the morphology of the star forming region. We use data on kinematics of molecular gas to support the hypothesis of induced star formation, and distinguish three basic types of molecular gas components. The first type is primordial undisturbed gas of the giant molecular cloud, the second type is gas entrained in motion by expansion of the \hii\ region (this is where the embedded clusters were formed), and the third type is a fast-moving gas, which might have been accelerated by winds from the newly formed clusters. The clumpy distribution of molecular gas and its kinematics around the \hii\ region implies that the picture of triggered star formation around S235 can be a mixture of at least two possibilities: the ``collect-and-collapse'' scenario and the compression of pre-existing dense clumps by the shock wave.
\end{abstract}


\section{Introduction}

A mechanism for sequential star formation, in particular the formation of OB subgroups in molecular clouds, was presented by \citet{ell_lad_77}. Since that time, it has been extensively studied both theoretically and numerically (see e.g. ~\citet{iau237} for recent results).
The essence of the triggered star formation is the dynamical influence of shock waves and winds from young massive stars and supernova explosions on their surroundings. The energy input leads to shock front propagation and dense shell formation around massive stars or their clusters. A shock front can compress pre-existing dense clumps forcing them to contract and form new stars. The swept-up shell itself may become gravitationally unstable and fragment into new protostars.

In their numerical study of star formation, triggered by expanding \hii\ regions, \citet{hosok_06} have argued that any expanding \hii\ region is able to trigger star formation if the ambient molecular material is dense enough. The similar conclusion has been reached by \citet{dale_07} on the basis of SPH modelling. However, these theoretical statements require an observational check, especially in cases where the environment of an \hii\ region possesses pronounced clumpiness and inhomogeneity.

In this paper we present a study of the molecular environment of the \hii\ region S235
(henceforth the S235 molecular complex). The complex is situated at a distance between 1.6~kpc and 2.5~kpc \citep{georg_73,isr_felli}, in the Perseus Spiral Arm. It appears to be a part of the giant molecular cloud G174+2.5, which has been observed in \dvco\ and \co\ lines by \citet{heyer_96}. The S235 region has been created by the massive O9.5V star BD~+35$^{\circ}$1201~\citep{georg_73}, which has ionised and dispersed the surrounding molecular gas. The giant molecular cloud G174+2.5 harbors 3~other optical \hii\ regions, namely, S231, S232, and S233. Shock waves, expanding away from these regions, should strongly influence surrounding molecular gas, changing significantly the kinematic structure of the cloud. The S235 molecular complex is a site of active star formation, and objects at various stages of this process are observed there. First, there are young stellar clusters around S235. The clusters have been detected via analysis of the Spitzer data in vicinity of S235 by \citet{allen_05} and via inspection of 2MASS images toward IRAS source 05382+3547 by \citet{kumar_06}. In the Section~\ref{res_counts} of this paper on the basis of more extensive star counts we confirm presence of these clusters and provide evidences for existence of some other infrared clusters in the region. Some members of these clusters are visible in POSS2/UKSTU Red and Infrared images taken from the Digitized Sky Survey archive. More deeply embedded and younger stellar/protostellar sources are present in the region: the source IRAS 05382+3547 was selected as a candidate precursor of ucHII region by \citet{molinari_91} on the basis of its colours. A 6.7~GHz methanol maser detected toward this infrared source by \citet{szym_00} clearly indicates presence of an embedded young massive star. Noteworthy, IRAS 05382+3547 is situated further away from S235 \hii\ region than the young stellar cluster \citep{kumar_06}. Our study shows that the separation between the center of the closest young cluster S235~East1 and the IRAS source is considerable and amounts to about \offsets\,=(--80\arcsec,50\arcsec). Young stellar sources associated with methanol masers are usually found to be members of stellar groups \citep{hunter_06,longmore_06}. Consequently, the existence of a much younger stellar cluster, not yet pronounced even in Spitzer images, can be inferred. The presence of various evolutionary stages suggests that star formation in this region can be a sequential process. 

In this study we consider star formation scenario in the vicinity of S235. Specifically, we examine if the expansion of S235 is responsible for the birth of young clusters around it. We found specific features of star formation triggered by the expansion of an \hii\ region, which were described in a series of papers of Deharveng, Zavagno et~al. \citep{deh_s104,deh_i,deh_ii}: 1) radio emission in molecular lines as well as IR continuum emission of PAHs and dust show that there is a shell-like structure of molecular material close to the edge of the \hii\ region and 2) positions of the young stellar clusters coincide with location of this gas shell. The procedure and results of star counts are described in Section~\ref{obs_counts} and Section~\ref{res_counts}, correspondingly. Observations of emission in the \cs\ and \co\ lines are summarised in Section~\ref{obs_emission}. The spatial distribution and the kinematics of molecular gas are described in Sections~\ref{res_molgas} and~\ref{gas_kin}. We investigated kinematics of the dense molecular gas (observed by \cs\ emission) and more rarefied gas (by \co\ emission) in details and contemplated that the observed kinematics might be one of the key instruments to distinguish between past, present and future stages of the triggered star formation process. This idea as well as the overall picture of star formation in the S235 molecular complex is presented in Section~\ref{disc}.

\section{Observational Data}\label{obs}

\subsection{Molecular Emission}\label{obs_emission}
We mapped the emission in the \cs\ and \co\ transitions toward the south-eastern part of S235 with the
Onsala Space Observatory 20-m telescope in December 2005 and February 2006.
The observations in December 2005 were dedicated to a search for peaks in emission in the \cs\ and \co\ lines around the 6.7~GHz methanol maser at $\alpha({\rm J2000.0}) = 05^{\rm h}41^{\rm m}33\fs 80^{s}$ and $\delta({\rm J2000.0}) = +35^{\circ}48^{'}27\farcs 00$. We did extended mapping and observed emission in \cvo\ and \ctrs\ transitions toward selected positions in February 2006. We spent about 27 hours of ``ON'' observational time for this project in February 2006. The main observational parameters are presented in Table~\ref{obs_par}. The frequencies of observed emission lines were taken from Lovas' catalogue~\citep{lovas}. The full width at half maximum (FWHM) beam-size at both frequencies was about 40\arcsec, and we used this value as a spacing for the \cs\ and \co\  maps. The main beam efficiencies were about 0.54 for 97~GHz and 0.50 for 110~GHz. We used a bandwidth of 40~MHz with 1600 frequency channels, which corresponds to a velocity range of about 110~\kms\ and a resolution of about~0.07~\kms. After the December 2005 session we found that the emission in \co\ and \cs\ is extended, and performed observations in a frequency-switch mode. The system temperature, including the atmosphere, was typically 300-500~K for observations of \cs\ and 600-800~K for \co. The typical noise level in maximum spectral resolution is 0.3~K for \cs\ and 0.8~K for \co.
The averaged uncertainties in the pointing accuracy were estimated to be 2.3\arcsec\ in azimuth and 4.2\arcsec\ in elevation.

\begin{table}
\caption{Basic observational parameters}
\begin{tabular}{ccccc}
\hline
Transition & Frequency  & FWHM      & $\Delta v$    & r.m.s.\\
           & (MHz)      & (\arcsec) & (\kms)        & (K)   \\
\hline
\co        & 110201.353 & 34        &   0.068       & 0.8   \\
\cs        & 97980.968  & 38        &   0.076       & 0.3   \\
\hline
\end{tabular}
\label{obs_par}
\end{table}

\subsection{Star Counts}\label{obs_counts}
In order to find young star clusters we calculated the surface density distribution of the sources from 2MASS Point Source Catalog. This catalogue contains all stars detected in at least one of 3 filters (J, H and K$_s$) used for the survey. Every star which appeared in the catalogue was used for the star counts. Having made star counts at infrared wavelengths, we are able to find clusters which are not yet seen at visible wavelengths. Values of stellar density were determined at a rectangular grid by the kernel
estimator method~\citep{silverman}. We performed counts with different angular scales. The first map is a large-scale distribution of stellar density for a field of 100\arcmin~$\times$~100\arcmin. A grid of 201~$\times$~201 nodes was used with a grid step of 0.5\arcmin. The kernel halfwidth was
chosen to be 1\arcmin. The second map was constructed for the vicinity of S235 with the dimensions of 20\arcmin~$\times$~20\arcmin.
We used a grid of 451~$\times$~451 nodes, a grid step of 0.2\arcmin, and a kernel halfwidth of 0.4\arcmin.
Results are presented in Section~\ref{res_counts}.

\section{Results}\label{res}

\subsection{Spatial distribution of molecular gas}\label{res_molgas}

\begin{figure*}
\includegraphics[scale=0.65]{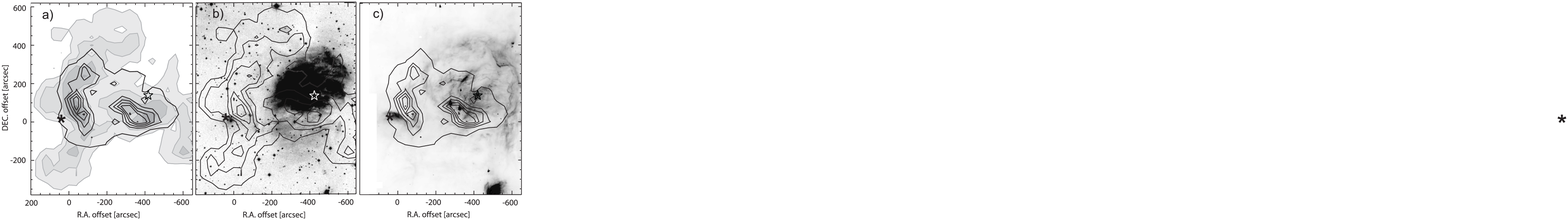}
\caption{ (a) Maps of \co\ (grey scale) and \cs\ (contours) emission toward S235. The contour levels for \cs\ map are shown from 1~K~\kms\ to 12~K~\kms\ with a step 3~K~\kms. The contour levels for \co\ map are shown from 5~K~\kms\ to 55~K~\kms\ with a step 10~K~\kms. Position of IRAS~05382+3547 and BD35$^{\circ}1201$ are shown by asterisk and starlet correspondingly. The origin of the maps is $\alpha({\rm J2000.0}) = 05^{\rm h}41^{\rm m}33\fs 80^{s}$ and $\delta({\rm J2000.0}) = +35^{\circ}48^{'}27\farcs 00$. (b) Optical image of S235 taken from DSS (POSS2 Red filter) and overlayed map of \co\ emission (contours) from S235. Origin of the plot and the contour levels are the same as in panel (a). Positions of IRAS~05382+3547 and BD35$^{\circ}1201$ are shown as in panel (a). (c) 8.0$\mu$m image of S325 from the GLIMPSE survey made with IRAC instrument from the Spitzer telescope archive and overlayed map of \cs\ emission (contours). Origin of the plot and the contour levels are the same as in panel (a). Positions of IRAS~05382+3547 and BD35$^{\circ}1201$ are shown as in panel (a).}
\label{int_13co_cs}
\end{figure*}

Maps of \cs\ and \co\ emission toward S235 are presented in Fig.~\ref{int_13co_cs}a. The origin of the map is $\alpha({\rm J2000.0}) = 05^{\rm h}41^{\rm m}33\fs 80^{s}$ and $\delta({\rm J2000.0}) = +35^{\circ}48^{'}27\farcs 00$. Multiple emission peaks are evident in both transitions. There is a peak in both transitions near (--40\arcsec, 40\arcsec), while western peaks  ($\Delta \alpha \approx - 400''$) do not coincide. Regions with the most intense \co\ and \cs\ emission form a shell-like structure. Comparison of the maps with the radio continuum image of S235 at $\lambda$21~cm
made by \citet{isr_felli} shows that the shell surrounds S235 at the south-eastern side. This is also clearly seen in Fig.~\ref{int_13co_cs}b, which is a composite of an optical image taken from the Digitized Sky Survey (POSS2/UKSTU Red) and the map of \co\ emission.
At the inner side, this molecular shell is bordered by the dark filaments (white in the negative image) located to the south and south-east of S235 (Fig.~\ref{int_13co_cs}b), which might well represent the dust collected by the shock wave. We suggest that the concentric dust and molecular shells have been shovelled by the expansion of the S235 \hii\ region. This suggestion will be further supported by consideration of the kinematics of the molecular and ionised gas.

In Fig.~\ref{int_13co_cs}c we show the map of \cs\ emission superimposed on the negative 8.0$\mu$m image of the molecular complex taken from the Spitzer Space Telescope archive. The mid-IR emission comes from the strong PAH features at 7.7 and 8.6$\mu$m~\citep{li_01}. It is clearly seen that the dense clumps are bordered by the mid-IR patches from the side facing S235. This infrared emission appears as a result of excitation of the PAHs in the thin layer at the border of the dusty region by the UV light of a close bright star or stellar cluster (e. g.~\citet{povich_07}).

\subsection{Young stellar clusters in vicinity of S235}\label{res_counts}

\begin{figure*}
\includegraphics[scale=0.9]{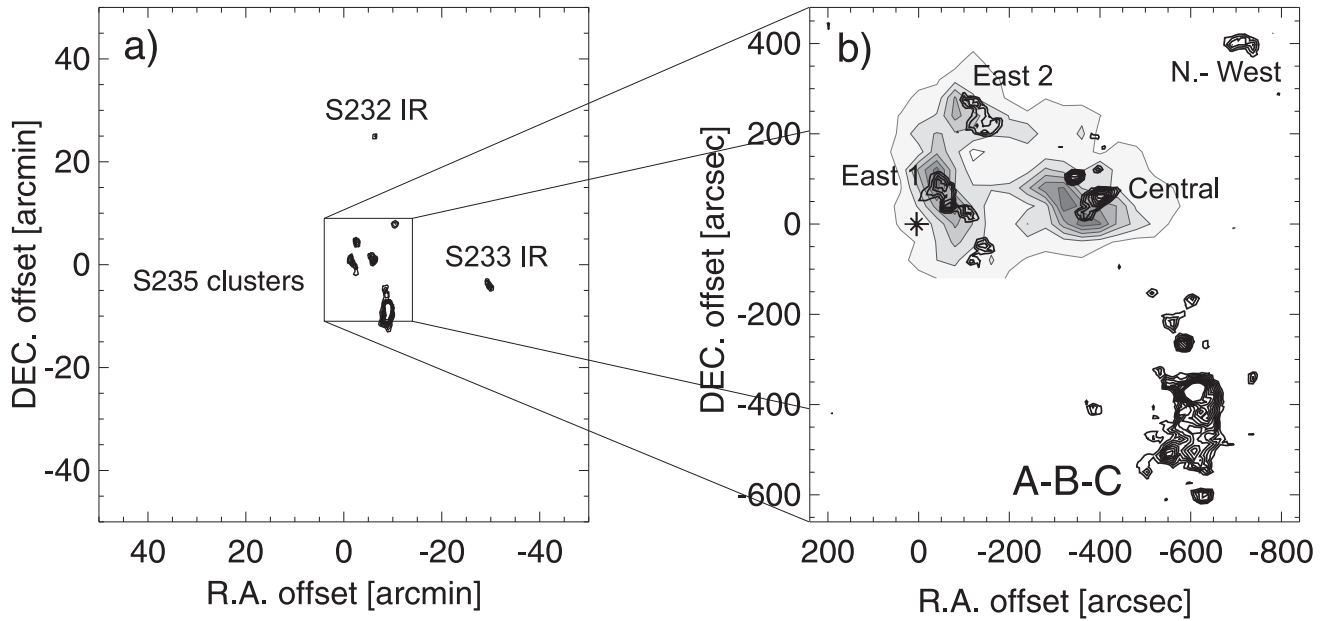}
\caption{a) Maps of the stellar density around S235. The origin of the plot is the same as in Fig.~\ref{int_13co_cs}.
Contours correspond to a stellar density range from 13~${\rm stars} \cdot {\rm arcmin}^{-2}$ to
31~${\rm stars} \cdot {\rm arcmin}^{-2}$ with a step 2~${\rm stars} \cdot {\rm arcmin}^{-2}$.
b) Map of CS(2-1) emission and stellar density distribution in the nearest vicinity of S235.
The origin of the plot is the same as in Fig.~\ref{int_13co_cs}. Contours correspond to a stellar
density range from 20~${\rm stars} \cdot {\rm arcmin}^{-2}$ to 56~${\rm stars} \cdot {\rm arcmin}^{-2}$ with a step of 4~${\rm stars} \cdot {\rm arcmin}^{-2}$. Weak variations of background are omitted. Position of IRAS~05382+3547 is shown by asterisk.}
\label{af_fig56}
\end{figure*}

We searched the environment of S235 for infrared star clusters to locate sites of recent star formation
in a much wider area than in previous studies~\citep{allen_05,kumar_06}.
The stellar density distribution in the region around S235 is shown in Fig.~\ref{af_fig56}.
Fig.~\ref{af_fig56}a shows the large-scale map (100\arcmin~$\times$~100\arcmin). Seven stellar clusters are found,
listed with their coordinates and designations in Table~\ref{clust_coord}.
The richest cluster in the map is associated with the well-studied star-forming region containing objects S235A, S235B and S235C \citep{felli_97}. We designate it S235~A-B-C.
Clusters  S235~A-B-C, S235~East1 and Central are shown in the figures by \citet{allen_05}.
Clusters S235~East1 and East2 were considered by \citet{kumar_06}.
Two other clusters are situated more than 20\arcmin\ away from S235: these are the quite well studied binary cluster S233IR
between S231 and S233 \citep{porras_00} and the relatively poor stellar
condensation at the western edge of S232. There is also a cluster associated with IRAS source 05361+3539 (G173.58+2.45) studied by \citet{shep_02} which is relatively poor and indistinguishable among random condensations of the 2MASS sources.

So, our analysis brought identification of 2 new clusters, S235~North-West and S232~IR (Fig.~\ref{af_fig56}b). These clusters are relatively poor and we did not map them in \cs\ and \co\ lines.

Despite the thorough analysis, no other mid-IR clusters have been found in the 100\arcmin~$\times$~100\arcmin\ area around S235. It is noteworthy that five of seven bright clusters are located around S235 within an area with a radius of about 10\arcmin. So, S235 seems to be a focus for young clusters in the whole giant molecular cloud.

\begin{table}
\caption{Coordinates of the young stellar clusters in the S235 molecular complex}
\begin{tabular}{lll}
\hline
Name & R.A. (J2000.0) & Dec. (J2000.0) \\
     & ($^h$ $^m$ $^s$) & ($^{\circ}$ \arcmin\ \arcsec)  \\
\hline
S235 East 1     & 5 41 29.55 &  35 49 21.65   \\
S235 East 2     & 5 41 24.19 & 35 52 15.96    \\
S235 Central    & 5 41 07.85 & 35 49 16.80    \\
S235 North-West & 5 40 45.73 & 35 55 05.72   \\
S232 IR         & 5 41 07.80 & 36 12 27.00    \\
S233 IR         & 5 39 10.95 & 35 45 18.83    \\
S235 A-B-C      & 5 40 51.94 & 35 41 35.49    \\
IRAS 05361+3539 & 5 40 54.45 & 35 44 17.39    \\
\hline
\end{tabular}
\label{clust_coord}
\end{table}

The close vicinity of S235 was investigated with finer resolution. Contours of stellar density distribution,
superimposed on the map of \cs\ emission, are shown in Fig.~\ref{af_fig56}b. It is clearly seen that
clusters S235~East1, East2, Central, North-West and S235A-B-C consist of sub-clusters and there are small
satellite groupings around them. The existence of sub-clusters was reported by \citet{aarseth} as a
feature inherent to the young age of clusters, which is supposed to be less than one dynamical time.
Our molecular line map covers the region containing three young clusters, S235~East1, East2 and Central.
The locations of \cs\ clumps correlate with these clusters, which confirms the extinction-based conclusion
that they are still embedded in dense clumps of parental material. The densest parts of each cluster are
projected onto the dense clumps. There is some displacement of \cs\ emission peaks relative to peaks of
the stellar density in the East2 and Central clusters which will be discussed in the next section.
So, our data shows that young stellar clusters, associated with dense clumps of parental material, are embedded in the molecular shell-like structure around the \hii\ region.

\subsection{Kinematics of molecular gas in the S235 complex}\label{gas_kin}

The locations of the young stellar clusters suggest that their formation might have been
triggered by expansion of S235. As the vicinity of S235 is quite clumpy, conclusions based on morphology should be corroborated by the studies of gas kinematics. Kinematics of ionised gas was considered by ~\citet{lafon_83}. He observed a velocity gradient from about --13~\kms\ to --30~\kms\ from the South-East to the North-West. The data is incomplete and does not allow to study kinematics of the ionized gas in necessary details. So, systemic velocity of the molecular gas of the giant molecular cloud G174+2.5 as a whole is more practical for fiducial point of the gas velocities. We discuss the velocity field of ionised gas in Sec.~\ref{disc} in available details. Our observations show that in vicinity of S235 the majority of the \co\ and \cs\ emission has velocities in the range from --25~\kms\ to --15~\kms\ which is blue-shifted with respect to the overall velocity distribution of the giant molecular cloud which peaks at about --17,5~\kms and has FWHP about 7~\kms\ (see Fig.5 of \citet{heyer_96}). Radial velocity of the ionising star BD~+35$^{\circ}$1201 is about --18~\kms\ as measured by the study of the photospheric absorption lines in the red optical range using 6-m BTA telescope at the Special Astrophysical Observatory of the RAS (Mlodik, Yushkin \& Sobolev, private communication). 

\begin{figure*}
\includegraphics[scale=0.99]{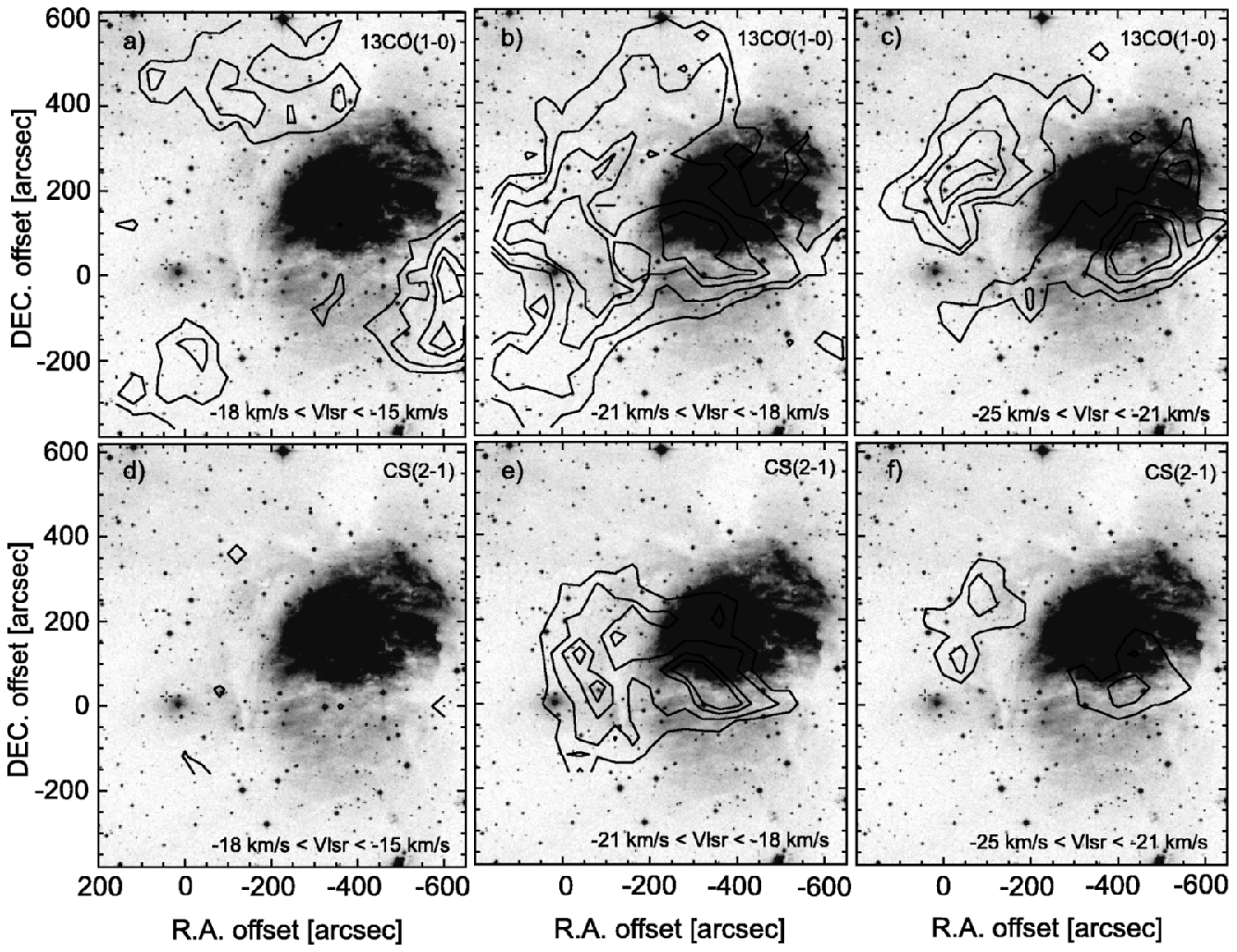}
\caption{Maps of \co\ (top row) and \cs\ (bottom row) integrated emission in different ranges of radial velocity superimposed on the optical image of S235 taken from the DSS. Origin of the plot is the same as in Fig.~\ref{int_13co_cs}. (a) and (d) --~18~\kms $<$ \vlsr $<$ --~15~\kms, (b) and (e) --21~\kms $<$ \vlsr $<$ --18~\kms, (c) and (f) --25~\kms $<$ \vlsr $<$ --21~\kms. Contour levels are given for 10, 25, 50 and 70\% from maximum values of the integrated intensity, which are 25 K~\kms\ for \co\ and 10 K~\kms\ for \cs\ emission.}
\label{13co_3comp}
\end{figure*}

 Analysis of our observational data reveals three distinct kinematical components. Spatial distributions of \co\ and \cs\ emission in velocity ranges from --15~\kms\ to --18~\kms\ (``red'' range), --18~\kms\ to --21~\kms\ (``central''), and --21~\kms\ to --25~\kms\ (``blue'') are shown in Fig.~\ref{13co_3comp}a and~\ref{13co_3comp}d, ~\ref{13co_3comp}b and~\ref{13co_3comp}e, \ref{13co_3comp}c and \ref{13co_3comp}f, respectively.

Gas responsible for emission in the ``red'' velocity range looks like three separate clumps visible in
the \co\ line and situated at the edges of the map. There is no \cs\ emission (tracing dense gas) in
the ``red'' range in the vicinity of S235, and there are no enhancements of stellar density toward these regions.
In the large-scale map of \co\ emission from the entire G174+2.5 molecular cloud \citep{heyer_96} the emission,
confined within the ``red'' velocity range, while also clumpy, appears to be distributed much more uniformly
than the \co\ emission in the other ranges of \vlsr. Thus, we suggest that the gas emitting in the ``red''
range in our map represents quiescent gas of the parental G174+2.5 molecular cloud, which was not disturbed by the expansion of S235 and can be considered as ``primordial'' with respect to the gas where the young stellar clusters have been formed.

\begin{figure*}
\includegraphics[scale=0.9]{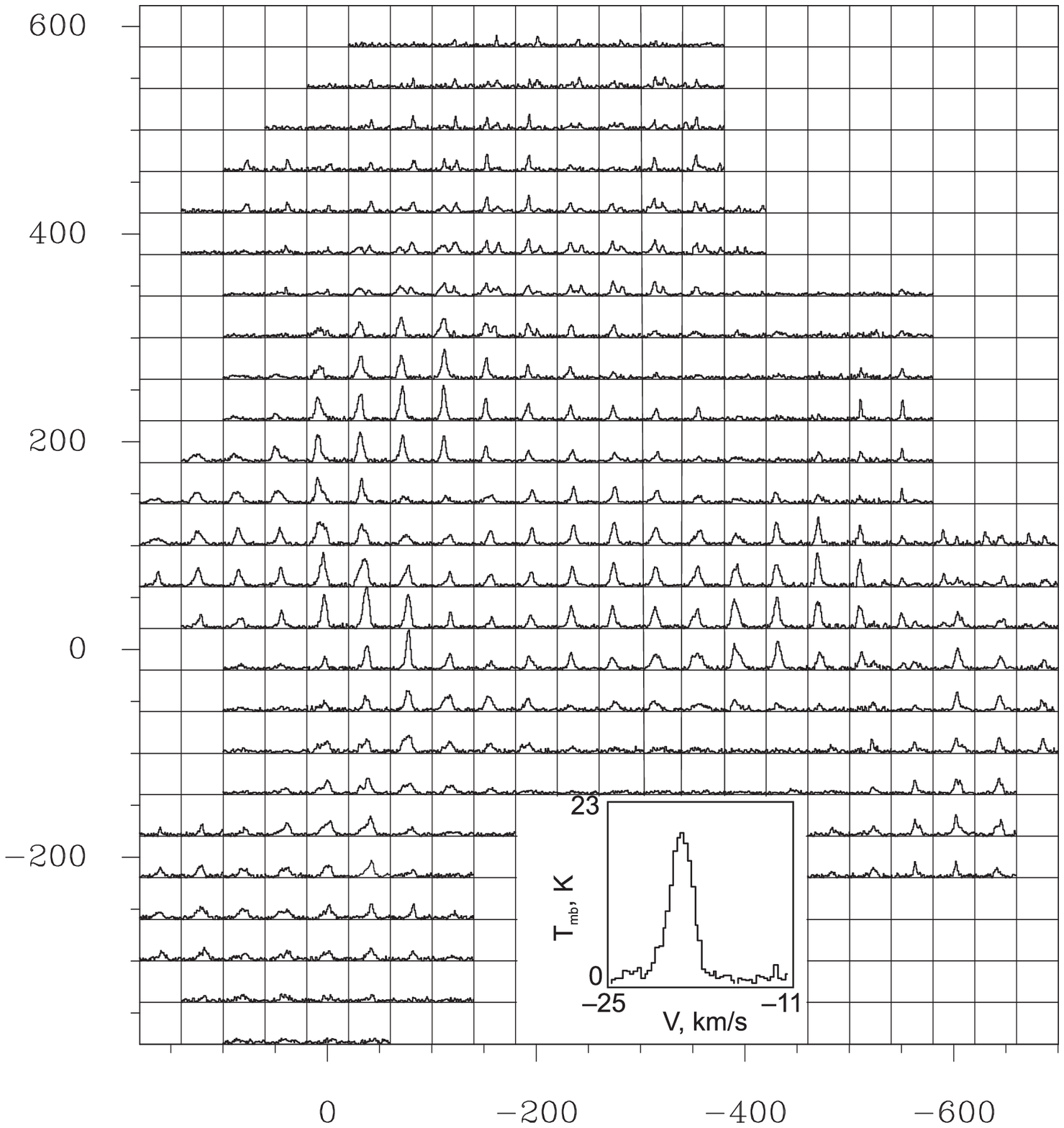}
\caption{Spectral map of \co\ emission. Origin of the map is the same as in Fig.~\ref{int_13co_cs}. Size of each cell equals 40\arcsec\,$\times$40\arcsec. The spectra are not smoothed. Scales of \vlsr\ and main beam temperature are shown in inset spectra (\offsets = (--80\arcsec, 40\arcsec)).}
\label{13co_spectra}
\end{figure*}

Emission in the ``central'' velocity range within the region, mapped in \co\ and \cs, is spread more uniformly
in our map in comparison with the other velocity ranges. The \co\ emission with the ``central'' velocities
is present almost in all spectra, except for the most south-western part of the map.
This is shown in Fig.~\ref{13co_spectra}, which contains spectra of the \co\ line toward all observed positions.
The transition from emission in predominantly ``central'' range to emission in predominantly ``red'' range is
seen toward  positions with double peaked \co\ lines.

Emission from dense gas in the \cs\ line is observed mostly in the ``central'' velocity range,
similarly to emission in \co\ line (see Fig.~\ref{13co_3comp}b and c). \cs\ emission is clearly associated
with the young stellar clusters (see Fig.~\ref{af_fig56}b). This is consistent with the assumption that the
``central'' velocity range represents gas, which has been affected and entrained in motion by the shock,
propagating away from S235, and, thus, attained the velocity difference.
Later, the young clusters were formed in the dense condensations of this gas.

\co\ and \cs\ emission in the ``blue'' range of \vlsr\ is confined to two well-defined patches which are located close to the S235~East2 and Central clusters. These are the most blue-shifted \vlsr\ values for the whole molecular cloud G174+2.5, and this emission is observed only towards these two young clusters. The well-defined location of the ``blue'' clumps and the noticeable difference between the features seen in the ``central'' and ``blue'' ranges lead to the assumption that the ``blue'' patches might represent dense (\cs) and surrounding more rarefied (\co) gas blown out by the stellar wind from the newly born clusters.

The young clusters might have started to emerge from the parental material, pushing out surrounding molecular gas.
The process was studied theoretically by, for example, ~\citet{weaver_77}.
This process might produce a measurable velocity difference between the dense gas moved by the stellar wind
and the surrounding rarefied gas, which has not yet been impacted by the wind.
Images of S235~East2 and Central clusters made in the K$_s$-band (2MASS Survey) and at 4.5$\mu$m (IRAC camera) contain filaments of dust which, according to their shapes, are most probably produced by the shock wave expanding from S235. Spectra of stars in the young clusters are needed to validate the idea about the origin of the ``blue'' component. Indication that the stars have strong enough winds would be a reliable confirmation.

\begin{figure}
\includegraphics[scale=0.33,angle=270]{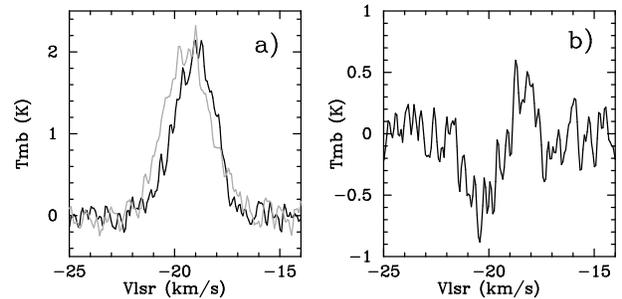}
\caption{Molecular spectra toward position with the offset coordinates (-40\arcsec, 40\arcsec). a) CS(2-1) (grey) and scaled \co(1-0) (black) spectra. b) Residuals of spectra shown in the panel a).}
\label{shift}
\end{figure}

Yet another kinematical evidence of the gas flows in the region is shown in Fig.~\ref{shift}.
Profiles of \co\ and \cs\ lines toward the embedded clusters demonstrate relative velocity shift.
The spectra toward (\offsets\,~=~(--40\arcsec , 40\arcsec)) have very good signal-to-noise ratio and
we illustrate the shift using this particular location (which is in the S235~East1 cluster).
We scaled the \co\ spectrum to the \cs\ line amplitude (see Fig.~\ref{shift}a)
and made a velocity channel-by-channel subtraction of the \cs\ spectrum from it (see Fig.~\ref{shift}b).
It is clear that the difference is large with respect to the noise level.
A possible explanation of this velocity shift is the following:
a wind blowing towards us compresses the gas
and imparts on it a more negative \vlsr\ with respect to the undisturbed gas, which is uncompressed and, hence,
relatively less dense. Since the \cs\ emission traces more dense (compressed) gas than the \co\ emission
(tracing undisturbed gas), we see the observed relative velocity shift.
Note that both the \co\ and \cs\ lines peaks inside the same (``central'') velocity range here. The non-detection of emission in the ``blue'' range toward the East1 cluster might be related to
a less evolved stage of the wind-driven outflow than in the East2 and Central clusters. Probably, the wind from the East1 cluster did not disperse the dense outflowing material yet, and we observe the velocity difference.

\section{Discussion}\label{disc}

\subsection{New view on star formation in the S235 molecular complex}

The possibility of triggered star formation in the S235 molecular complex by expansion of the \hii\ region has not been considered before.
We found evidence supporting our view on star formation in this region in the \ha\ line emission data
around S235 presented by \citet{lafon_83}. They found a gradient of \vlsr(\ha) from \vlsr~$>$~--15~\kms\ in the southern part to \vlsr~$<$~--25~\kms\ in the northern part of S235 where molecular gas is almost absent. Our map of \co\ emission does not contain data on this part of the S235 complex, but channel maps of \dvco\ and \co\ emission in this region can be found in \citet{heyer_96}. The channel maps allow suggesting that the northern part of the ionised nebula is expanding freely toward the observer. Data on radial velocities of radio recombination lines H109$\alpha$ by \citet{kazes}, on H167$\alpha$ and H140$\alpha$ by \citet{silver} and on H91$\alpha$ by \citet{quireza} toward some positions in S235 are in agreement with the gradient found by \citet{lafon_83}.

The kinematic pattern of molecular gas was interpreted by \citet{evans_81} (observations
of \dvco, \co, \cvo, H$_2$CO(2$_{21}$-1$_{11}$), HCN(1--0), HCO$^+$(1--0) emission)
and \citet{lafon_83} (HCO$^+$(1--0) and \co\  transitions) as a signature of existence of two molecular clouds with the mean radial velocities of --17~\kms\ and --20~\kms. Molecular gas to the south of S235, which contains the star forming regions S235A, S235B, and S235C, was interpreted as the ``--17~\kms'' cloud and gas to the east of S235, containing young clusters S235~East1 and East2, as the ``--20~\kms'' cloud. \cite{evans_81} mentioned that it is difficult to distinguish between the two clouds in directions where both velocity components are present in their map, but a thorough analysis of molecular kinematics was not done.

Values of \vlsr(\ha) around --17~\kms\ are observed in the direction of the southern part of the \hii\ region which
is shielded by molecular gas around the S235 Central cluster. It is easy to see in Fig.~\ref{13co_3comp}
that the \vlsr\ of \co\ emission there is mainly in the range from --18~\kms\ to --21~\kms, and absent at --17~\kms.
It seems that molecular gas in front of the \hii\ region moves toward us faster than the ionised gas.
This is expected if the gas is pushed by the shock wave. It was shown by numerical modeling by \citet{hosok_06}
and \citet{hosok_07} that the velocity difference between the shock and ionisation fronts can be
several kilometers per second if an \hii\ region is surrounded by uniformly distributed gas.
The difference becomes smaller if the gas density falls off with the distance from the \hii\
region as $n \sim r^{-w}$. It was shown by \citet{franco_90} that the critical value of $w$,
when the initial ionised region still has a finite size, depends on a ratio of model parameters describing
expansion of the \hii\ regions in a non-uniform medium, $w_c \sim (1-(r_c/R_S)^3)^{-1}$, where $r_c$
is a radius of inner flat part in the density distribution and $R_S$ is the Str\"omgren radius.
We see that the molecular gas affected by the shock front (i.e., emitting in the ``central'' velocity range)
propagates faster than the ionisation front, which means that the initial gas density distribution was
not steep enough for formation of an unbound \hii\ region. Numerical modelling of the coupled chemistry
and dynamics of an \hii\ region, expanding into a clumpy medium with regions of decreasing and, probably,
increasing gas density, is needed to clarify the history of star formation in S235 molecular complex.

The star formation in S235A-B-C is likely to be initiated by a
process different from the expansion of S235. These regions are
situated quite far from the border of the S235, which is
delineated in the radio continuum observations by
\citet{isr_felli}. Studies of these regions by \citet{nakano_86}
and \citet{felli_97} showed that the bulk of dense gas emission in the
optically thin \cvo\ and \ctrs\ transitions appear in the \vlsr\
range from about --18~\kms\ to --15~\kms. This indicates that
these star-forming regions are likely embedded in primordial gas
of the G174+2.5 giant molecular cloud.

\subsection{Scenario of star formation in the S235 molecular complex}

There are several known ways for an \hii\ region to trigger star formation around itself
(see \citet{elm_98} for a review). In particular, contraction of pre-existing molecular clumps can be stimulated by the shock wave propagating away from the \hii\ region. This process was investigated, e.g., by \cite{boss_95} for shocks with velocities of about 10~\kms\ and low-mass molecular clouds, and by \cite{krebs_83} for rapid shocks (faster than 200~\kms) and high-mass clouds. This scenario predicts random distribution of dense clumps with embedded stellar clusters around an \hii\ region. Ionised boundary layers \citep[similar to observed, e.g. by][]{thom_04} are expected at the sides of the clumps which face the ionising star.

Another way is the formation of star-forming cores via ``collect-and-collapse'' scenario \citep{ell_lad_77}. It predicts the existence of young clusters embedded in the shell of molecular gas located ahead of the ionisation front. In an ideal spherically-symmetric case this gas is expected to be observed as a ring of molecular emission around an \hii\ region. There should also be rings of mid-IR emission produced by PAHs in the PDR region and millimetre dust emission around the \hii\ region. The young clusters have to be situated on a neutral side of the compressed layer, between the PDR and the molecular gas of the parental cloud.

There are three morphological signatures of the ``collect-and-collapse'' scenario here: 1) young stellar clusters are embedded in the dense molecular shell around S235, 2) PAHs residing in the PDR (see Fig.~\ref{int_13co_cs}c for emission at 8.0$\mu$m) produce the shell-like mid-IR emission, 3) the young clusters S235~East1 and East2 are situated on the neutral side of the mid-IR shell. The mid-IR shell looks broken toward the S235~Central cluster, which is projected on the optical image of the \hii\ region, but East1 and East2 are located to the east of S235. All three clusters are embedded in clumps emitting in the ``central'' velocity range. This fact is an evidence supporting ongoing ``collect-and-collapse'' scenario.

It is not easy to single out one scenario of triggering of star formation by S235 because the molecular environment of the \hii\ region is clumpy. The actual scenario can be a mixture of the scenarios mentioned before. The answer can be prompted by consideration of evolutionary stages of the embedded clusters. The age similarity between the three clusters would rather be an argument in favour of a ``collect-and-collapse'' scenario, as it seems less probable to find three independent pre-existing clumps with similar mass and density on a similar distance from the \hii\ region. On the other hand, even in a ``collect-and-collapse'' scenario, clusters do not need to be exactly coeval, the age distribution might be caused by different densities of initial clumps in a fragmenting molecular shell. We discuss evolutionary stages of the embedded clusters below.

\subsection{Evolution stages of the young embedded stellar clusters}

Our data on the distribution of the stellar density and the dense gas as well as the kinematic data indicate that the clusters S235~East2 and Central are probably more evolved than East1. Peaks of the stellar density are displaced relative to the peaks of \cs\ emission for the
S235~Central and East2 clusters. These clusters might have already started active dispersion of the dense gas by their stellar winds, giving rise to the ``blue'' component of the \co\ and \cs\ clumps in Fig.~\ref{13co_3comp}c. Observations of Herbig Ae/Be stars by~\citet{fuente_98,fuente_02} showed a sequence of objects which represents a history of mass dispersal around them: from embedded Herbig Ae/Be stars to stars located in a cavity within the molecular cloud. They suggested that energetic winds from the stars swept out gas and dust, forming a cavity. Similar phenomenon can take place in S235.

In contrast, the S235~East1 cluster does not show noticeable displacement of the molecular emission peak relatively to the stellar density maximum and does not show any associated ``blue'' range emission. This cluster might be younger. Alternatively, S235~East1 can be at the same evolutionary stage as the other two clusters, if it does not contain stars which are luminous enough to produce a fast and powerful impact on its environment. Data on \cvo\ emission toward the clusters are needed to estimate masses of the three dense clumps and those of the gas emitting in the ``blue'' velocity range.

The question about relative ages of the clusters could be answered by spectral studies of their stellar population. \citet{allen_05} identified some of young stellar objects, belonging to the S235~East1, East2 and Central clusters, as Class~I and Class~II sources. They found Class~I objects in all three clusters, but Class~II objects were discovered only in the S235~East2 and Central clusters. This result is in agreement with our guesses described above. But this result cannot be considered as a strong evidence of age spread because of small number of identified objects (two young stars in S235~East1, seven in S235~East2 and 10 in S235~Central). More objects must be measured to make definite conclusions.

It is tempting to suggest that kinematic components, which we see in S235, correspond to sequential evolutionary stages. In this line of reasoning, the most advanced stage of the triggered star formation would be represented by the East2 and Central clusters, where a non-negligible amount of gas has already been accelerated by stellar winds. The clump coincident with the East1 cluster would be at the intermediate stage as it is actively forming stars but have not yet developed any observable wind signatures.

Interesting is also the clumpy ``red'' component, constituting the structure of the parent molecular cloud. We do not know for sure if the three ``clumps'' seen on the \co\ map are not just projection effects, but real physical entities. But if they are real, they may represent a visible part of a more abundant population of clumps, waiting for some external influence to trigger the formation of star clusters and massive stars in them. As such, these clumps are of particular interest, and we plan to continue the observational study of the S235 complex, with the ultimate goal to check if we do see objects in this area which represent three distinct stages related to past (S235~East2 and Central), present (East1), and future (hypothetical ``red'' clumps) of the triggered formation of star clusters and massive stars.

\section{Conclusions}\label{conc}

We investigated the kinematics and distribution of molecular gas and the spatial distribution
of young stellar clusters in the S235 molecular complex and found evidences that the process of star formation in the complex was triggered by expansion of the \hii\ region.

The \hii\ region S235 is surrounded by a shell-like envelope with dense molecular clumps at the eastern and
southern sides. The shell circles the ionisation front and is separated from the \hii\ region by a PDR envelope
visible in mid-IR. The dense molecular clumps contain young stellar clusters not yet seen at visible
wavelengths. We suggest that the expansion of S235 is responsible for the formation of the clusters.

Emission from molecular gas in the \co\ and \cs\ transitions reveals three kinematically distinct components, which can be interpreted in terms of the triggered star formation scenario, presumably representing three stages in the influence of young stellar objects on the surrounding gas. The clumpy gas emitting in the ``red'' velocity range (--18\kms\,$<$~\vlsr~$<$--15\kms) may represent primordial material of the G174+2.5 giant molecular cloud undisturbed by expansion of S235. This gas is not dense enough to be readily observable in \cs\ emission. Emission in the ``central'' range (--21~\kms~$<$~\vlsr~$<$--18~\kms) can be produced by the gas clumps which were affected and entrained in motion by the expansion of S235. These are the clumps where the young clusters were formed. Removal of molecular gas by the stellar winds from the clusters S235~East2 and Central might have given rise to the ``blue'' component (--25 \kms\,$<$~\vlsr~$<$--21 \kms) in \co\ and \cs\ emission.

The cluster S235~East1 is probably younger and represents an earlier evolutionary stage than the Central and East2 clusters. Another possibility is that S235~East1 contains less luminous stars, being at a similar evolutionary stage. More extensive spectral data on the stellar population of these clusters are needed to determine their relative ages.

\section*{Acknowledgments}

MK and DW thank for financial support the RFBR (grant 07-02-01031-a) and the INTAS Foundation (YS PhD Fellowship, grant 05-109-4862). AMS was supported by the RFBR grant 07-02-00628-a. MK thanks the staff of the Onsala Space Observatory for the hospitality and assistance during observations in 2005. A considerable part of molecular line observations for this project was performed by Lars E.B Johansson who passed away on March 15, 2008. He is deeply missed by us. We are grateful to A.P. Tsivilev for providing us with the references regarding recombination radiolines observations. Special thanks to Mlodik and Yushkin who allowed us to show the data on the BD~+35$^{\circ}$1201 velocity prior to publication. We thank the anonymous referee for critical comments and suggestions which helped us to clarify our statements and improve the presentation. Special thanks to Paul Boley for corrections of English in some parts of this paper. 

This study has made use of the Digitized Sky Survey, produced at the Space Telescope Science Institute, and the Spitzer Telescope Archive.

\label{lastpage}

\end{document}